\documentclass[11pt,reqno,a4paper,twoside]{article}

\usepackage{amsmath,amsthm,amstext,amscd,amssymb,euscript,mathrsfs}
\usepackage{calrsfs}
\usepackage{epsfig}

\usepackage{color}
\long\def\red#1{{\color{black}#1}}

\newcommand{\Z}{\mathbb Z}
\newcommand{\R}{\mathbb R}
\newcommand{\N}{\mathbb N}

\newcommand{\E}{\mathbb E}

\renewcommand{\Pr}{\mathbb P}

\renewcommand{\phi}{\varphi}
\newcommand{\epsi}{\ensuremath{\epsilon}}

\newcommand{\pee}{\ensuremath{\mathbb{P}}}

\newcommand{\loc}{\mathcal{L}}

\def\1{{\mathchoice {\rm 1\mskip-4mu l} {\rm 1\mskip-4mu l}
{\rm 1\mskip-4.5mu l} {\rm 1\mskip-5mu l}}}

\newtheorem{theorem}{{\small T}{\scriptsize HEOREM}}[section]
\newtheorem{corollary}{{\bf{\small C}{\scriptsize OROLLARY}}}[section]
\newtheorem{proposition}{{\bf{\small P}{\scriptsize ROPOSITION}}}[section]
\newtheorem{lemma}{{\bf{\small L}{\scriptsize EMMA}}}[section]
\newtheorem{remark}{{\bf{\small R}{\scriptsize EMARK}}}[section]
\newtheorem{definition}{{\bf{\small D}{\scriptsize EFINITION}}}[section]

\renewenvironment{proof}[1]
{\noindent{{\bf{{\small P}{\scriptsize ROOF}}}.}\hspace{0.1cm} #1} {$\;\qed$\newline}

\newcommand{\beq}{\begin{eqnarray}}
\newcommand{\eeq}{\end{eqnarray}}

\newcommand{\ba}{\begin{align*}}
\newcommand{\ea}{\end{align*}}

\newcommand{\be}{\begin{equation}}
\newcommand{\ee}{\end{equation}}

\newcommand{\bl}{\begin{lemma}}
\newcommand{\el}{\end{lemma}}

\newcommand{\br}{\begin{remark}}
\newcommand{\er}{\end{remark}}

\newcommand{\bt}{\begin{theorem}}
\newcommand{\et}{\end{theorem}}

\newcommand{\bd}{\begin{definition}}
\newcommand{\ed}{\end{definition}}

\newcommand{\bp}{\begin{proposition}}
\newcommand{\ep}{\end{proposition}}

\newcommand{\bc}{\begin{corollary}}
\newcommand{\ec}{\end{corollary}}

\newcommand{\bpr}{\begin{proof}}
\newcommand{\epr}{\end{proof}}

\newcommand{\bi}{\begin{itemize}}
\newcommand{\ei}{\end{itemize}}

\newcommand{\ben}{\begin{enumerate}}
\newcommand{\een}{\end{enumerate}}

\newcommand{\caA}{{\mathcal A}}

\newcommand{\caC}{{\mathscr C}}

\newcommand{\caH}{{\mathcal H}}

\newcommand{\caK}{{\mathcal K}}
\newcommand{\caL}{{\mathcal L}}

\newcommand{\caP}{{\mathcal P}}

\newcommand{\caR}{{\mathcal R}}

\begin{document}
\title{Dynamics of condensation in the symmetric inclusion process}
\author{ Stefan Grosskinsky$^{\textup{{\tiny(a)}}}$,
Frank Redig$^{\textup{{\tiny(b)}}}$, Kiamars Vafayi $^{\textup{{\tiny(c)}}}$\\
{\small $^{\textup{(a)}}$
Mathematics Institute,}\\
{\small University of Warwick}\\
{\small Coventry CV4 7AL, UK}\\
{\small $^{\textup{(b)}}$
Delft Institute of Applied Mathematics,}\\
{\small Technische Universiteit Delft}\\
{\small Mekelweg 4, 2628 CD Delft, Nederland}\\
{\small $^{\textup{(c)}}$ Department of Mathematics and Computer Sciences}\\
{\small Technische Universiteit Eindhoven,}\\
{\small Postbus 513,
5600 MB Eindhoven, Nederland}
}

\maketitle

\begin{abstract}

The inclusion process is a stochastic lattice gas, which is a natural bosonic counterpart of the well-studied exclusion process and has strong connections to models of heat conduction and applications in population genetics. Like the zero-range process, due to attractive interaction between the particles,
the inclusion process can exhibit a condensation transition.  In this paper we present first rigorous results on the dynamics of the condensate formation for this class of models. We study the symmetric inclusion process on a finite set $S$ with total number of particles $N$ in the regime of strong interaction, i.e. with independent diffusion rate $m=m_N \to 0$. For the case $Nm_N\to\infty$ we show that on the time scale $1/m_N$ condensates emerge from general homogeneous initial conditions, and we precisely characterize their limiting dynamics.
In the simplest case of two sites or a fully connected underlying random walk kernel, there is a single condensate hopping over $S$ as a continuous-time random walk. In the non fully connected case several condensates can coexist and exchange mass via intermediate sites in an interesting coarsening process, which consists of a mixture of a diffusive motion and a jump process, until a single condensate is formed. 
Our result is based on a general two-scale form of the generator, with a fast-scale neutral Wright-Fisher diffusion and a slow-scale deterministic motion. The motion of the condensates is described in terms of the generator of the deterministic motion and the harmonic projection corresponding to the absorbing states of the Wright-Fisher diffusion.

\bigskip

\noindent

\end{abstract}

\section{Introduction}
The inclusion process, introduced in \cite{gkr} as the dual of a model of heat conduction, and further developed in \cite{grv},
is a natural counterpart of the well-known and extensively studied exclusion process.
The inclusion process on a finite set $S$ combines random walk jumps at rate $m\, p(i,j)/2$ with inclusion jumps,
where 
each particle at site $i$ is attracted by each particle at $j$ independently with rate $p(i,j)$.
Here $p(i,j), i,j\in S$ are the rates of a symmetric irreducible
continuous-time random walk on $S$. The inclusion jumps introduce a form of attractive interaction between the particles.

When $m$ tends to
zero and simultaneously many particles are in the system, due to the attractive interaction between the particles, large piles of particles
will be formed at individual sites.
In fact we proved in \cite{grv1} condensation for the stationary measure,
in a limit where both
$m\to 0$ and the number of particles $N\to\infty$.
In this paper we prove dynamical results for the condensation phenomenon. More precisely, we study
how clusters arise from an arbitrary initial condition, and
how they move and merge into a single condensate, which then 
jumps over the finite lattice as a random walk.
We therefore make $m=m_N$ dependent on $N\in \N$ such that $m_N\to 0$ as $N\to\infty$, which implies
that diffusion will get slower and the attractive inclusion interaction will create condensates.

This model can be interpreted as a multi-allele version of the Moran model \cite{moran}, describing the evolutionary competition of several species in a fixed size population. The inclusion part describes reproduction and death and $m_N$ plays the role of an additional mutation rate, which is typically very small on the reproduction time scale (see e.g. \cite{arnoldt} and references therein). It is also important as a dual process to models of energy or momentum transport \cite{gkr,grv,gkrv,ber}.

Related recent theoretical results include
explosive condensation in a totally asymmetric model \cite{waclaw} which exhibits a
slinky motion of the condensate also observed recently in \cite{gunter} for non-Markovian zero-range dynamics. Results on the equilibration dynamics in zero-range processes with decreasing rates are currently under investigation \cite{jara}.
In order to have a well defined limit dynamics we require symmetry of the $p(i,j)$, but our proof is not based on potential theoretic methods as used in \cite{beltran,landim} for zero-range processes. Due to the system-size dependence of the diffusion rate we can apply more general multi-scale methods (see e.g. \cite{stuart}). In particular, we use a two-scale decomposition of the generator
and construct the generator of the limit process similarly to results in \cite{kurtz}.

The decomposition consists of a Wright-Fisher type diffusion part which runs at ``infinite speed''
(in the limit $N\to\infty$), and a $p(i,j)$-dependent drift part. The limiting motion is then described
by the harmonic projection of the drift part on the absorbing set of the diffusion.
In the simplest case of two sites, or similarly, in the fully connected case where all $p(i,j)$ are strictly positive,
the absorbing set of the diffusion is the set of corner points of the simplex
$ E= \{ x\in [0,1]^S: \sum_{i\in S} x_i=1\}$. This corresponds to the immediate formation of a single
condensate, which then hops over the set $S$ as a random walk. In that case we can characterize the limit dynamics in a relatively simply way as the solution to a martingale problem for linear functions.
In the more general case, in a first stage, several condensates will form, and interact via intermediate sites.
Condensates do not split but can merge, and this coarsening dynamics
eventually leads to a single condensate which then again moves over $S$ as a random walk.
In this case, we compute the generator explicitly making use of the separation of time scales and the martingales for the fast Wright-Fisher diffusion.

The rest of our paper is organized as follows. In Section \ref{sec:res} we introduce the model and present the main results and possible extensions. In Section \ref{sec:lemma} we collect general results on tightness and the comparison of semigroups and work out the simple two-site and fully connected cases. The proof of the general case is treated in Sections \ref{sec:gen} and Sections \ref{sec:final} where we give a general definition of the limiting generator of a two-scale system in terms of projection operators on harmonic functions. In Section \ref{sec:threesite} we give an explicit computation of the limiting generator in the case of 3 lattice sites, which already contains most of the difficulties of the general case.

\section{Definitions and main results\label{sec:res}}

\subsection{The model and the auxiliary slow-fast system}

We consider a finite set $S$, with associated
jump rates $p(i,j)=p(j,i)$, $j,i\in S$ of a symmetric, irreducible random walk \red{with $p(i,i)=0$.}

The symmetric inclusion process $\big(\eta (t):t\geq 0\big)$ with parameter $m>0$
based on $p(i,j)$
is then defined  as in \cite{grv} to be the continuous-time Markov process on the configuration
space $\Z_+^S$, with generator
\be\label{sipgen}
\caL^{SIP}_N f(\eta )= \sum_{i,j\in S} p(i,j)\eta_i\left(\frac{m}{2}+\eta_j \right) \left(f(\eta^{ij})-f(\eta)\right)\ ,
\ee
where $\eta_i \in\Z_+$ denotes the number of particles at site $i\in S$, and $\eta^{ij}_k =\eta_k -\delta_{i,k} +\delta_{j,k}$, $k\in S$ denotes the configuration where one particle moved from $i$ to $j$. 
Since the total number of particles
is conserved, we consider the process with generator \eqref{sipgen} on the state space
$\Omega_N =\big\{\eta\in\Z_+^S :\sum_{i\in S} \eta_i =N\big\}$ with a fixed total number $N\in \N$ of particles.

We consider $m=m_N$ dependent on the number of particles $N$ such that $m_N\to 0$ as $N\to\infty$, which implies
that diffusion becomes slower and the attractive inclusion interaction will create
condensates. We are interested in the limiting dynamics of the rescaled process after accelerating time by a factor $\theta_N := \alpha /m_N$ with $\alpha >0$. We will assume throughout the paper that
\be\label{mainas}
m_N \to 0\ ,\ \ N m_N \to\infty \quad\mbox{so that}\quad \frac{\theta_N }{N}=\frac{\alpha}{m_N N}\to 0\ .
\ee

We denote by $e_i, i\in S$ the canonical unit vectors
of $l_1(S)$ with entries $e_i (j) = \delta_{i,j}$.
Consider the rescaled process $\big( x^N (t):t\geq 0\big)$ defined by
\[
x^N (t):=\big(\eta_i (\theta_N t)/N: i\in S\big).
\]
This is a Markov process on the simplex $E=\big\{ x\in [0,1]^S :\sum_{i\in S} x_i =1\big\}$ with generator
\be\label{xgener}
\caL_N f(x) = \sum_{i,j\in S}\theta_N p(i,j)Nx_i \left( \tfrac{m}{2} + Nx_j\right) \Big( f( x{-}\tfrac{1}{N} e_i {+}\tfrac{1}{N} e_j)- f(x)\Big)\ .
\ee

Assuming smooth $f$ in \eqref{xgener}, Taylor expansion of the right-hand side 
gives, using the symmetry of $p(i,j)$,
\beq\label{taylor1}
\caL_N f(x)&=& -\sum_{i,j\in S}\frac{\alpha}{4}p(i,j) (x_i-x_j) (\partial_{ij} f) (x)
\nonumber\\
& &+ \frac12\sum_{i,j\in S}p(i,j) x_i x_j \theta_N (\partial_{ij}^2 f)(x) + O(\theta_N /N)\ ,
\eeq
where we abbreviated
\be\label{abbr}
\partial_{ij}:=\left(\frac{\partial}{\partial x_i}-\frac{\partial}{\partial x_j}\right)\ .
\ee
Here, by \eqref{mainas}, the correction term denoted by $O(\theta_N /N)$ is a function of $x$ which converges to zero uniformly in $x$, as $N\to\infty$. We will show that
this correction terms can be ignored and one can study the easier auxiliary process $\big( y^N (t):t\geq 0\big)$ on $E$ with generator
\beq\label{taylorgen}
L_N&:=& L +\theta_N L'\nonumber\\
&=& -\sum_{i,j\in S}\frac{\alpha}{4}p(i,j) (y_i-y_j) \partial_{ij}
+ \theta_N \frac12\sum_{i,j\in S}p(i,j) y_i y_j \partial_{ij}^2 \ .
\eeq
As $N\to\infty$ this is a slow-fast system with two scales, where the fast part
\be\label{difgenpartL'}
L'=
\frac12\sum_{i,j\in S}p(i,j) y_i y_j \partial_{ij}^2
\ee
corresponds to a Wright-Fisher diffusion with absorbing states given by 
\be\label{abss}
\caA:=\big\{ x\in E\, :\, p(i,j)x_i x_j =0\mbox{ for all }i,j\in S\big\}\ .
\ee
The slow part is a deterministic motion with generator
\be\label{detgenpartL}
L=-\sum_{i,j\in S}\frac{\alpha}{4}p(i,j) (y_i-y_j) \partial_{ij} \ .
\ee

The limiting dynamics of the process with generator $L_N$
is intuitively described as follows. The fast ($\theta_N L'$)-part of the generator pushes
``infinitely'' fast to the set of absorbing
states $\caA$, whereas the slow ($L$) part tries to move away from $\caA$. So the motion
consists of infinitesimally moving away from $\caA$ and immediately being
projected back onto $\caA$. This will result in an ``effective, projected process'' on the set $\caA$.
We will describe this procedure of ``projection'' rigorously and in more detail in section \ref{sec:projection} below.

\subsection{Main results}

Our main results rigorously characterize the limiting dynamics
of the process $(x^N(t):t\geq 0)$, by showing 
convergence to a limit process $\big( x(t):t\geq 0\big)$ which concentrates on the subset $\caA\subset E$.
Furthermore, the set of corner points
\be\label{corner}
\caC :=\big\{ e_i :i\in S\big\}\subset \caA
\ee
is in turn absorbing for the limit process, as we can conclude from the first theorem.
In the following we speak of weak convergence on path space, if for every $T>0$ the processes $(x^N(t), 0\leq t\leq T)$ converge to $(x(t),0\leq t\leq T)$ weakly in the Skorokhod topology.

\bt\label{theo2}
\ben
\item
Assume that $x^N (0)\to x^0\in \caC$ in distribution. Then $\big( x^N (t):t\geq 0\big)$ converges weakly on path space to $\big( x(t):t\geq 0\big)$, which is a Markov chain on $\caC$ with initial condition $x(0)=x^0$ and generator
\be\label{cornergencase}
Af(e_i)=\sum_{j\in S} \frac{\alpha}{2} p(i,j)\big( f(e_j )-f(e_i)\big)\ .
\ee
\item
In the fully connected case, i.e., if $p(i,j)>0$ for all $i,j\in S$, then the same holds for a general initial condition. I.e., if $x^N (0)\to x^0\in E$ in distribution, then $\big( x^N (t):t\geq 0\big)$ converges weakly on path space to $\big( x(t):t\geq 0\big)$, the process on $\caC$ with generator \eqref{cornergencase}, and
with initial condition $\pee (x(0)=e_i)= \E [x^0_i ]$.
\een
\et

So if the process is started from a configuration in $\caC$, i.e.,  when (as $N\to\infty$) all the mass concentrates on a single site, all future configurations are of this type and the single pile of mass (condensate) performs a random walk on $S$ with rates proportional to $p(i,j)$. The same holds in the fully connected case for general initial conditions, since in that case $\caA =\caC$ and only the corner points are absorbing for the Wright-Fisher part. For general symmetric $p(i,j)$, $\caA$ has a more complicated structure and the limit process is not ergodic on $\caA$. In the following we will describe the dynamics for general initial conditions under the additional assumption
\be\label{ass01}
p(i,j)\in \{0,1\}\quad\mbox{for all }i,j\in S\ ,
\ee
i.e., where edges are either connected or not with uniform jump rate $1$. The more general case can be treated as well but is more complicated to formulate, and we will comment on that in a discussion in Section \ref{sec:diss}. Let
\be\label{nnn}
\hat p(i,j)=\big( 1-p(i,j)\big)\sum_{k\in S} p(i,k)p(k,j)\geq 0
\ee
be the number of two-step connections between sites $i,j\in S$, which are not directly connected.

To state the result we also need to introduce the harmonic measure $\nu_x$ on $\caA$ for the Wright-Fisher diffusion with generator $L'$ given in (\ref{difgenpartL'}), with $x\in E$ as initial condition.
This is the distribution of the point $x(\infty)\in \caA$ where the diffusion with generator $L'$ is absorbed when
started from $x\in E$ (see Section \ref{sec:projection} below for more details). For the initial condition in Theorem \ref{theo2} item 2, note that for non-random $x^0$ the harmonic measure is exactly $\nu_{x^0} (e_i )=x_i^0$, since the marginal dynamics on each site is given by a diffusion (see Section \ref{sec:full} for details) and only the corner points are absorbing. For the general case below there is no easy explicit formula for $\nu_{x^0}$.

\bt\label{theo3}
Assume (\ref{ass01}) and assume that at time zero $x^N (0)\to x^0\in E$ in distribution. Then $\big( x^N (t):t>0\big)$ converges weakly on path space to $\big( x(t):t\geq 0\big)$, which is a Markov process on $\caA$ with initial condition $x(0)\sim \E [\nu_{x^0} ]$
and generator
\beq\label{lima}
\lefteqn{Af(x)=\sum_{i,j\in S} \frac{\alpha}{2}\hat p(i,j) x_i x_j 
\partial_{ij}^2 f(x)}\nonumber\\
& &+\! \sum_{j\in S}\! \delta_{x_j ,0} \left(\sum_{i\in S}\frac{\alpha}{2} p(i,j) x_i \right)\!
\red{\left( f\Big( x{+}\sum_{i\in S}p(i,j) x_i \big( e_j {-}e_i\big)\Big) {-}f(x)\right) .}
\eeq
\et
{\
The generator of this process consists of two parts. The first part corresponds to an effective Wright-Fisher diffusion between sites which are connected by a path of length two only, which results from exchanging mass via the intermediate site. This mechanism of mass exchange is also studied heuristically in \cite{waclaw} for a totally asymmetric model. The second part corresponds to all the neighbouring mass of an empty site $j$ accumulating on that site with a rate proportional to the total neighbouring mass. In the simplest case where there is only one neighbour with non-zero mass, this corresponds to a jump of a single pile. Therefore piles can exchange mass continuously or merge, but they never split and the set of corner points $\caC$ where all mass concentrates in a single pile is absorbing. For the inclusion process this absorbing set is reached on the same time scale $\theta_N$ as the stationary motion on the corner set takes place. This is different from other models such as the zero-range process, 
where these dynamics happen on different time scales \cite{grossketal03,godreche,beltran,landim,jara}.

Note that the initial condition for the limit process is not $x^0$ and convergence in the usual sense does not hold at time $0$, since the process is  ``immediately'' projected onto the absorbing set $\caA$. The initial condition is then distributed as the absorption point on $\caA$ starting from $x^0$, which is consistent with the right limit $x(0+)$.

\subsection{Discussion\label{sec:diss}}

\begin{figure}
\begin{center}
\includegraphics[width=0.8\textwidth]{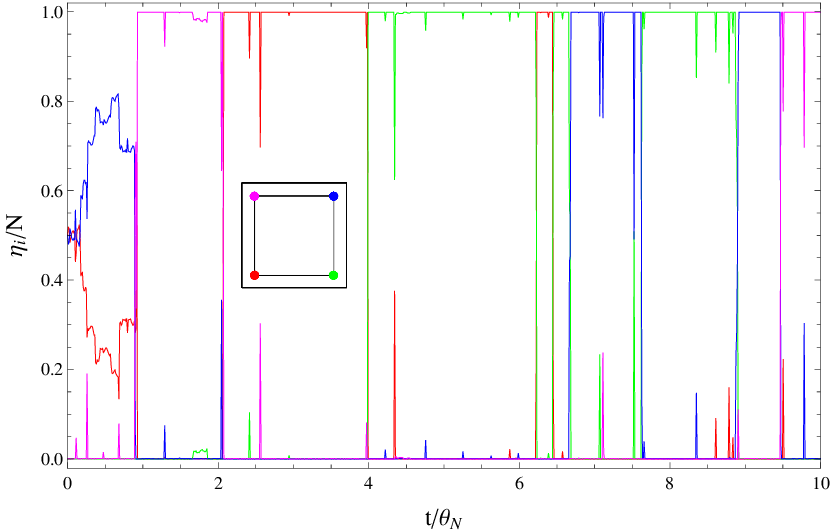}
\end{center}
\caption{\label{fig1}
Simulation illustrating the main results by a sample path of the process with normalized occupation numbers \red{$\eta_i /N$ shown in different colors}. The lattice is $S=\{ 1,\ldots ,4\}$ with nearest neighbour connections and periodic boundary conditions. Initially, there is diffusive mass exchange between sites which are not directly connected, and later all mass concentrates on a single pile. Fluctuations are due to finite parameter values, where $N=1000$, $m_N =0.01$ and $\theta_N =100$.
}
\end{figure}

The general limit dynamics of Theorem \ref{theo3} is illustrated in Figure \ref{fig1} for a ring of 4 sites. After initial diffusive mass exchange between sites 1 and 3, the system reaches the absorbing set of corner points and turns into a jump process. Figure \ref{fig2} illustrates the fully connected case (second item in Theorem \ref{theo2}) where the system instantaneously reaches a corner and a single pile performs a random walk on the lattice.

\begin{figure}
\begin{center}
\mbox{\includegraphics[width=0.48\textwidth]{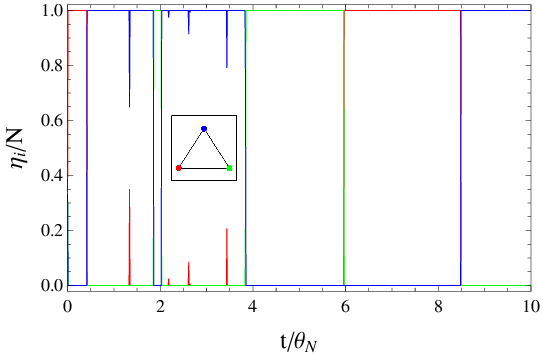}\quad\includegraphics[width=0.48\textwidth]{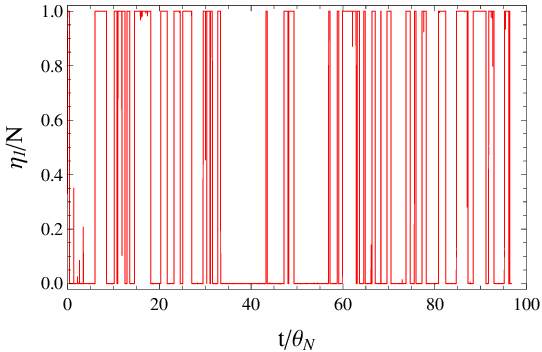}}
\end{center}
\caption{\label{fig2}
Simulation illustrating the main results for a fully connected lattice $S=\{ 1,2,3\}$ of three sites, where immediately all mass concentrates on a single pile. The left panel shows the evolution of the occupation number of all sites in different colors, and the right panel the evolution of site $1$ for longer times. Fluctuations are smaller than in Fig.~\ref{fig1} since here $N=10000$, $m_N =0.001$ and $\theta_N =1000$.
}
\end{figure}

Our results hold for symmetric jump rates $p(i,j)$, since otherwise the Taylor expansion (\ref{taylor1}) would contain first order drift terms diverging with $N$, which would lead to infinite limiting speed of the clusters. Therefore the limit dynamics are not well defined unless the $p(i,j)$ are symmetric.

In the formulation of Theorem \ref{theo3} we restrict ourselves to $p(i,j)\in \{ 0,1\}$, only for simplicity of presentation. Our methods can be applied straightforwardly to the general case, only the concrete computations in Section \ref{sec:threesite} have to be adapted. Assuming $p(1,2)=p(2,1)=p$, $p(2,3)=p(3,2)=q$ and $p(1,3)=p(3,1)=0$ for the case of 3 sites, the continuous part of the limit generator $A$ (\ref{lima}) then takes the form
\[
\frac{\alpha}{4}(q{-}p)x_1 x_3 \,\partial_{13}+\frac{\alpha}{2} \big( qx_1 {+}(p{-}q)x_1^2\big)\big( px_3 {+}(q{-}p)x_3^2\big)\,\partial_{13}^2 \ .
\]
Compared to (\ref{apart2}) and (\ref{lima}), there is an additional drift term and the diffusive term is no longer of Wright-Fisher type. This can be formulated for general lattices analogously. 
The jump part of $A$ still takes the same form as in (\ref{lima}) on a general lattice.

\section{A technical lemma and first results for the two-site and the fully connected case\label{sec:lemma}}

This section consists of two parts.
\ben
\item  In section 3.1, we show the following.
If the auxiliary process $(y^N(t):t\geq 0)$ with generator $L_N$ (\ref{taylorgen})
converges to a limiting process $( y(t):t\geq 0)$  and if the ``true process'' $(x^N(t):t \geq 0)$ with generator $\loc_N$ \eqref{xgener}
converges along a subsequence, then this limit is the same process $(y(t):t\geq 0)$.
As a consequence, if the sequence $(x^N(t): t\geq 0)$ is tight, then
it has the limit characterized by the auxiliary process. This observation will be proved in Lemma \ref{auxlem} below and allows us to work with the simpler generator $L_N$ to characterize the limit process.
\item In subsection \ref{sec:twosite} below we first consider the case of two sites, which is easier than
the general case because the limiting process is a pure jump process on the set
of unit vectors $\{e_1, e_2\}$.
Then we work out the fully connected case, which is similar to the two site
case, and where the limiting process is also a pure jump process on the set of unit vectors
$\{e_i:i\in S\}$.
\een

\subsection{General results on comparison and tightness}

We start with a technical lemma that we will apply later to justify dropping the $O(\theta_N /N)$ part in the Taylor expansion of the generator (\ref{taylor1}).

\bl\label{auxlem}
Suppose $\loc_N$ and $L_N$ are generators of Markov processes
on the compact state space $E$, such that on a common core of functions $\caK\subset C(E)$ \red{contained in continuous functions $C(E)$} we have
\be\label{genas}
\|\loc_N f -L_N f\|_\infty \leq \||f\|| \beta_N \ .
\ee
Here $\| .\|_\infty$ denotes the supremum norm, $\beta_N$ is a sequence of numbers converging to zero as $N\to\infty$, and
$\|| .\||$ denotes a norm such that for all $f\in \caK$ and $t>0$
\be\label{supest}
\sup_N\sup_{0\leq s\leq t} \|| e^{sL_N} f\||<\infty\ .
\ee
Then we have:
\ben
\item For the associated semigroups, for all $t\geq 0$ and $f\in\caK$
\be\label{semico}
\lim_{N\to\infty}\|e^{tL_N} f- e^{t\loc_N} f\|_\infty =0\ .
\ee
\item
If $\lim_{N\to\infty}e^{tL_N}f =S_t f$ for all $t\geq 0$, $f\in C(E)$ with $(S_t :t\geq 0)$ a Feller semigroup, then
$\lim_{N\to\infty}e^{t\caL_N}f= S_t f$ as well.
\item
If the processes with generators $\loc_N$ and $L_N$ are tight on the path space $D(0,\infty )$ of rcll functions (right-continuous with left limits), then they have the same limit process.
\end{enumerate}
\el

\bpr
Pick $f\in \caK$.
Start from integration by parts (or Dyson's) formula (see e.g.\ \cite{liggett} p.~367)  and use that $e^{s\caL_N}$ is a contraction in the supremum norm to estimate
\beq
\|e^{tL_N}f-e^{t\loc_N}f \|_\infty &=& \Big\|\int_0^t e^{s\caL_N} (\loc_N -L_N ) e^{(t-s)L_N} f\, ds\Big\|_\infty \nonumber\\
&\leq &
\int_0^t \| (\loc_N -L_N ) e^{(t-s)L_N} f\|_\infty ds\nonumber\\
&\leq &
\int_0^t \beta_N \||e^{(t-s)L_N} f\|| ds\nonumber\\
&\leq &\beta_N t \sup_{0\leq s\leq t}\||e^{sL_N} f\|| \to 0
\eeq
as $N\to\infty$ for all $t>0$.

For the second item, let $f\in C(E)$, and for given $\epsi>0$ choose $g\in \caK$ such that $\|g-f\|_\infty <\epsi$. Then
for $N$ large enough we have, by item 1,
$\|e^{tL_N}g - e^{t\loc_N} g\|_\infty <\epsi$ and by the contraction property of $e^{tL_N}, e^{t\loc_N}$,
$\|e^{tL_N} (f-g)\|_\infty\, ,\ \|e^{t\loc_N } (f-g)\|_\infty <\epsi$.
This implies $\|e^{tL_N}f - e^{t\loc_N} f\|_\infty <3\epsi$ and if one of the terms converges to a semigroup, so does the other.

Finally, the third item follows because a Markov process on the path space $D(0,\infty )$ is uniquely determined by its semigroup.
\epr

\bl\label{tcrit}
The sequence of Markov processes $\big( x^N (t):t>0\big)$ on a complete separable metric space $E$ is tight on the path space $D(0,\infty )$ of rcll functions if
\begin{enumerate}
\item The sequence $\big( x^N (t):t\geq 0\big)$ is stochastically bounded on $D[0,\infty )$, and
\item For all $f\in C^\infty_c (E)$, denoting compact smooth functions (which are uniformly dense in $C(E)$) and all $\epsilon >0$ there exist non-negative random variables $Z_N (\delta, f,\epsilon )$ such that for all $\delta$
\be
\Big|\E \big[ f(x^N (t+u))-f(x^N (t))\big| x^N (t)=x\big]\Big|\leq \E\big[ Z_N (\delta, f,\epsilon )\big| x^N (t)=x\big]
\ee
with probability $1$ for all $t\geq\epsilon$ and $0\leq u\leq \delta$, and
\be
\lim_{\delta\searrow 0} \limsup_{N\to\infty} \E\big[ Z_N (\delta, f,\epsilon )\big] =0\ .
\ee
\end{enumerate}
\el

The conditions are sufficient but not necessary, for a proof, see e.g. \cite{whitt}, Lemma 3.11 or \cite{ethierkurtz}, Chapter 3. The first condition will always be fulfilled since our state space $E$ is a compact, finite dimensional simplex and $\| x^N (t)\|_\infty \leq 1$. The second condition has been modified for our purposes, and ensures boundedness of the modulus of continuity of sample paths starting the process from any strictly positive time and will be verified later. At time $0$ the sequence of processes we consider is actually not tight, since in the limit it gets instantly projected onto an absorbing state of the Wright-Fisher diffusion $L'$. So we can show convergence only for all positive times $t>0$, and by right-continuity of sample paths we can extend the limit process uniquely to all $t\geq 0$. We will show in Section \ref{sec:gen} that this extension is consistent with the harmonic measure, describing the absorption of the fast Wright-Fisher diffusion.

\subsection{The two site case\label{sec:twosite}}

We start now with the simplest case
\be\label{scase}
S=\{1,2\}\quad\mbox{and}\quad p(1,2)=p(2,1)=1\ ,
\ee
where $S$ contains only two sites and $\caA =\caC =\{ e_1 ,e_2\}$.

\bt\label{condpos2sites}
Assume (\ref{scase}) with initial conditions $x^N (0)$ such that $x^N (0)\to x^0 \in [0,1]$ in distribution as $N\to\infty$. Then the process $\big( x^N (t):t> 0\big)$ converges weakly on path space to $\big( x(t):t\geq 0\big)$, which is a simple random walk on $\{ e_1 ,e_2\}$ with jump rates $\alpha /2$ and initial condition $x(0)\in\{ e_1 ,e_2\}$ with $\Pr (x(0)=e_1 )=x^0_1$.
\et

The rest of this section is devoted to the proof. Since $S$ consists of two elements
we abbreviate $x_1=x$, and because in the process $x_1+x_2=1$ is conserved, we effectively
have only one variable (denoted by $x$). The
Taylor expansion leading to \eqref{taylor1} then gives
in terms of $x$:
\beq\label{tota}
\caL_N f(x) &=& \left(-\tfrac12\theta_N m_N (2x-1)\partial_x + 2 \theta_N x(1-x)\partial_x^2\right) f(x) + O(\theta_N /N)
\nonumber\\
&=&: L_N f(x) + O(\theta_N /N)
\eeq
where the terms $O(\theta_N /N)$ go to zero uniformly in $x$, as $N\to\infty$.
The generator
\be\label{farfala}
L_N= -\tfrac12\alpha (2x-1)\partial_x + 2\,\theta_N x(1-x)\partial_x^2
\ee
appearing in \eqref{tota} is a Wright-Fisher diffusion
with mutation where in front of the diffusion term $x(1-x)\partial_x^2$
we have a factor
$\theta_N$, i.e., the diffusion part (also called ``genetic drift part'' in the population dynamics language) of the
process is accelerated
w.r.t.\ the mutation process.

This implies intuitively that the process will tend to evolve
``immediately'' towards the fixed points
of the genetic drift process, which are the homo-zygotes $x=1$ and $x=0$, and the mutation process
will then lead to a random flipping between these two states.

First, we show that we can apply Lemma \ref{auxlem} to our case.
Since this is slightly technical, we state it as a separate lemma.
\bl
The generators $\caL_N$ and $L_N$ of \eqref{tota} satisfy the assumptions
of Lemma \ref{auxlem} with $\|| f\||= (\| f''\|_\infty +\|f'''\|_\infty )$ and
with $\caK$ the set of polynomials as a common core.
\el
\bpr
The approximate generator reads (for simplicity we put $\alpha /2=1$ in the following)
\[
L_N=-(2x-1)\partial_x + 2\,\theta_N x(1-x) \partial^2_x
\]
and for the difference with the true generator we have the estimate
\[
\|\caL_N f- L_N f\|_\infty \leq C\Big(\frac1N\| f''\|_\infty +\frac{\theta_N}{N} \|f'''\|_\infty \Big)\leq \frac{C\theta_N}{N}\big(\|f''\|_\infty +\|f'''\|_\infty \big)
\]
from a standard expression of the remainder in Taylor expansions. The norm
$\||f\||=(\|f''\|_\infty +\|f'''\|_\infty )$ can thus be used in \eqref{genas}.

So in order to see that \eqref{supest} holds for this choice of norm, it is sufficient to prove that for the auxiliary process $y^N (t)$ with generator $L_N$, the expectations
\[
\E_y f(y^N (s))=e^{sL_N} f(y)
\]
have bounded second and third derivatives in $x$, uniformly
in $0\leq s\leq t$ and $N\in \N$, for sufficiently many $f$.
Choosing for $f$ the polynomial $f_n (y)= y^n$, we have,
\[
L_N f_n = 2\,\theta_N n(n-1) (f_{n-1}- f_n) + n( f_{n-1}-f_n)-nf_n
\]
This can be interpreted as follows: the first two terms form the generator
of a Markov chain $(n_t :t\geq 0)$ that jumps from $n$ to $n-1$ at rate $2\,\theta_N n(n-1) + n$
corresponding to the classical ancestral dual of Wright-Fisher diffusion (Kingman's coalescent + mutation contribution), and
the last term is an extra killing part. For more details on that connection see e.g.\ the nice account on duality in population models \cite{huillet}. Using the Feynman-Kac formula we then obtain,
\[
\psi(N,t,n,y):=\E_y f_n(y^N (t))= \E_n^{N} \big( e^{-\int_0^t n_s ds} f_{n_t}(y)\big)
\]
where $\E_n^N$ denotes expectation for the process on $\N$ with generator
\[
K_N g(n)= \theta_N n(n-1) (g(n-1)-g(n)) +  n (g(n-1)-g(n))\ .
\]
As a consequence, since the process with generator $K_N$ only lowers the starting
$n$ in the course of time, we obtain the uniform estimate
\[
\Big|\frac{d^k}{dx^k} \psi(N,t,n,y)\Big| = \Big|\E_n^{N} (e^{-\int_0^t n_s ds} y^{n_t-k}n_t (n_t-1)\ldots (n_t-k+1))\big|\leq n^k
\]
for all $y\in [0,1]$, which gives \eqref{supest}. Finally, it is clear that on the compact simplex $E$ the polynomials form a common core of the generators $L_N, \loc_N$.
\epr

So Lemma \ref{auxlem} guarantees that we can identify the limit from the auxiliary process $\big( y^N (t):t\geq 0\big)$ with generator $L_N$.
We first show that the limit process concentrates on the set $\{ 0,1\}$ and then characterize it as a solution to the martingale problem
of the claimed limiting process.

\bl\label{one}
We have for all $t>0$
\be\label{cl1}
\sup_{y\in [0,1]}\E_y \big[ y^N (t) (1-y^N (t))\big]\to 0\quad\mbox{as }N\to\infty\ ,
\ee
and furthermore
\be\label{cl2}
\limsup_{N\to\infty} \theta_N \sup_{y\in [0,1]}\E_y \big[ y^N (t) (1-y^N (t))\big] \leq C
\ee
for some $C>0$. The same holds for $x^N (t) (1-x^N (t))$ of the original process with generator $\caL_N$.
\el

\bpr
We abbreviate $f(y)= 2y(1-y)$, $g(y)= 2(2y-1)^2$ and compute
\[
\frac{d}{dt} \E_y f(y^N (t))= \E_y (g(y^N (t)))  -2\,\theta_N\E_y f(y^N (t))\ .
\]
Putting $\psi_t (y)= \E_y f(y^N (t))$, $\phi_t (y)=\E_x g(y^N (t))$, we rewrite this equation as
\[
\frac{d}{dt} \psi_t = \phi_t-2\,\theta_N \psi_t \ ,
\]
which has the standard form of a slow-fast system (see e.g. \cite{stuart} for more background), omitting the fixed argument $y$. This can be rewritten as
\be\label{intresu}
\psi_t = \psi_0 e^{-2\theta_N t} + \frac{e^{-2\theta_N t}}{\theta_N} \int_0^{\theta_N t} \phi_{s/\theta_N} e^{2s} ds\ .
\ee
Using now the fact that $y\in [0,1]$, we obtain that $\phi \leq 2$ and $\psi\leq 1$ which gives
\be\label{clue}
0\leq \psi_t (y) \leq  e^{-\theta_N t}+ \frac{2}{\theta_N}\left(1- e^{-2\,\theta_N t}\right)
\ee
which goes to zero as $N\to\infty$ for all $t>0$, thus proving (\ref{cl1}). The estimate (\ref{clue}) also implies the second statement (\ref{cl2}).\\
The same argument holds for $x^N (t) (1-x^N (t))$ if we replace $\phi_t$ by $\phi_t +O(\theta_N /N)$ including the remainder terms, since we only use boundedness of $\phi_t$ for $N$ large enough.
\epr

This shows that the limit process (if it exists) concentrates on the state space $\{ 0,1\}$ for all $t>0$, and the estimate can also be used to show tightness using the criterion of Lemma \ref{tcrit}, which implies existence of a subsequential limit by Prohorov's theorem (see e.g. \cite{ethierkurtz}, Chapter 3).\\

\bl\label{ttight}
The processes $\big( y^N (t):t>0 \big)$ and $\big( x^N (t):t>0 \big)$ are tight on the path space $D(0,\infty )$.
\el

\bpr
As mentioned before, the state space $E=[0,1]$ is compact and the first condition of Lemma \ref{tcrit} holds clearly. For the second condition, we first note that the process is \textit{not} tight for $t=0$ since it is immediately projected onto the absorbing set $\caA$ which is $\{ 0,1\}$ in this simple case. For all $t>0$ we can show tightness for the auxiliary process first by noting that for any smooth function $f$ on $[0,1]$
\red{
\be\label{tbbounded}
\E \big[ f(y^N (t{+}u)){-}f(y^N (t))\big| y^N (t){=}y\big] =\E \Big[\int_t^{t+u} \!\!\!\! L_N f(y^N (s))\, ds\Big| y^N (t){=}y\Big]\ .
\ee
On the right-hand side 
\be
L_N f(y) =-(2y-1)f'(y)+\theta_N y(1-y)f''(y)
\ee
and the first term is bounded by $\| f'\|_\infty$ since $y\in [0,1]$. For the second term we can exchange expectation and time integration since the integrand is positive and bounded above, and use (\ref{intresu}) to get
\begin{eqnarray}
\lefteqn{\theta_N \,\E \left[\int_t^{t+u} y^N (s)\big( 1-y^N (s)\big)\, ds\Big|\, y^N (t){=}y\right]}\nonumber\\
&=&\int_t^{t+u}\left( \theta_N y(1{-}y)e^{-2\theta_N (s-t)} {+} e^{-2\theta_N (s-t)} \int_0^{\theta_N (s-t)} \!\!\!\!\!\phi_{w/\theta_N} e^{2w} dw\right)\, ds \nonumber\\
&\leq &\frac12 y(1-y) \big( 1-e^{-2\theta_N u}\big) +2u\ ,
\end{eqnarray}
where we also used boundedness of $\phi$ as before. This is independent of $t$ and for all $t\geq\epsilon$ and $0\leq u\leq\delta$ we can therefore use
\be
Z_N (\delta, f,\epsilon ):=\| f'\|_\infty +\Big( \tfrac12 y( 1-y) +2\delta\Big) \| f''\|_\infty
\ee
to bound the modulus of (\ref{tbbounded}) as in Lemma \ref{tcrit}. 
For every fixed $\epsilon >0$ this bound holds uniformly in $N$, and uniformly in the conditional value $y\in E$ at time $t$. Using that $y=y^N (t)$ for some $t\geq\epsilon$ it follows from Lemma \ref{one} that for all $y_0 \in E$
\be
\E_{y_0} \big[ Z_N (\delta, f,\epsilon )\big]\to 0
\ee
as $N\to\infty$ and $\delta\searrow 0$, since $f$ is smooth and all derivatives are bounded on $[0,1]$. This implies tightness by Lemma \ref{tcrit}. Since $\epsilon >0$ was arbitrary we have tightness for all strictly positive times $t>0$.}\\
The same argument holds for the original process $(x^N (t):t>0)$, since also third derivatives of $f$ are bounded and additional terms 
are of the form $\delta\| f'''\|_\infty (C+\theta_N /N)$.
\epr

In order to identify the limit process by a martingale problem, the following \red{well-known result} gives a simplified version for a process that jumps with rate $\lambda$ between $0$ and $1$.

\bl\label{two}
Let $X_t$ be a Markov process with values in $\{0,1\}$ and $\lambda >0$. If
\be\label{mar}
M_t :=X_t-X_0-\int_0^t \lambda(2X_t -1)ds 
\ee
is a martingale, then $X_t$ is the process that jumps with rate $\lambda$ between $1$ and $0$.
\el

\bpr
Denote by
\[
L f(x) =\lambda x (f(0){-}f(1)) + \lambda(1-x) (f(1){-}f(0))=-\lambda (2x-1) (f(1){-}f(0))
\]
the generator of the process
jumping back and forth between $0,1$ at rate $\lambda$.
Let $f: \{0,1\}\to\R$, then
$f(x)= xf(1) + (1-x) f(0)= x (f(1)-f(0)) + f(0)$ is always linear. Therefore, using \eqref{mar}
\be\label{pro}
f(X_t)-f(X_0)-\int_0^t Lf(X_s)= M^f_t
\ee
is a martingale for all $f:\{0,1\}\to\R$, which characterizes the process.
\epr

Now we can conclude the proof of Theorem \ref{condpos2sites}. Applying the generator $L_N$
\eqref{tota} of the auxiliary process to the function $f(y)=y$ gives that
\[
M^N (t):=y^N (t)-y^N (0)-\frac{\alpha}{2}\int_0^t \big( 2y^N (s)-1\big)\, ds
\]
is a martingale.
Therefore, if 
$\big( y^N (t):t>0\big)\to \big( x(t): t>0\big)$ along a subsequence,
for the limiting process we have that
\[
M(t) :=x(t)-x(0)-\frac{\alpha}{2}\int_0^t \big( 2x(s)-1\big)\, ds
\]
is a martingale. Thus the limit is the claimed jump process on $\{ 0,1\}$. By Lemma \ref{auxlem} this is also the limit of the original process $x^N (t)$.

Since the limit process $\big( x(t):t>0\big)$ has right-continuous paths the consistent initial condition is given by
\be\label{inicon}
x(0):=\lim_{\epsilon\searrow 0}\lim_{N\to\infty} y^N (\epsilon )\ ,
\ee
which concentrates on $\{ 0,1\}$ by Lemma \ref{one}. To see that this has the distribution claimed in Theorem \ref{condpos2sites}, denote $\phi(t,N)= \E_{y^N (0)} (y^N(t))$. Then
using the generator
\eqref{farfala}, we see that
\[
\frac{d \phi(t,N)}{dt} = -\red{\frac{\alpha}{2}}(2\phi(t,N)-1)
\]
which gives
\[
\phi(t,N)= y^N (0) + \frac12 (1-e^{-\red{\alpha} t})\ .
\]
Taking the limits $N\to\infty$ and then $t\downarrow 0$, and using that the limiting process is concentrated on $\{0,1\}$ gives the initial condition as claimed.

\br
We will see in Section \ref{sec:gen} that in general the limiting semigroup $S(t)$ has the property that $S(0)=P$ is a projection onto harmonic functions on $E$. So if $\mu$ is the distribution of limiting initial condition $x^0$ as given in Theorems \ref{theo2} and \ref{theo3}, the initial distribution of $y(0)$ for the limit process is given by $\mu P =\E_\mu [ \nu_{x^0} ]$ since $(Pf)(x)=\E_{\nu_x} [f]$ for all $f\in C(E,\R )$. For a symmetric diffusion on $[0,1]$ the harmonic measure $\nu_x$ on $\{ 0,1\}$ is then given by $\nu_x =x^0_1$, i.e. equal to the first component of $x^0 \in E$ (see the second item of Theorem \ref{theo2}).
\er

\subsection{The fully connected case\label{sec:full}}

We will see in the next section that Lemma \ref{auxlem}, the concentration Lemma \ref{one} and the tightness argument in Lemma \ref{ttight} can be extended straightforwardly to the general case. In case the set $S$ is fully connected, i.e. all $p(i,j)>0$, we have $\caA =\caC$ and the limit process still concentrates on the corner points and can be characterized by a simple martingale problem. This will provide a proof of the second statement of Theorem \ref{theo2}.

The following lemma gives the martingale characterization of the process
jumping with rates $p(i,j)$ from $e_i$ to $e_j$.

\bl\label{fclem}
Let $\big( y(t):t\geq 0\big)$ be a process on $\{e_i, i\in S\}$, and denote $y_i (t) = y(t).e_i= \1 (y(t)=e_i)$. If for all $i$
\be
M^i (t):=y_i (t)-y_i (0)-\sum_{j\in S}\int_0^t \left(y_j (s)-y_i (s)\right)p(i,j) ds 
\ee
is a martingale, then $y$ is the random walk jumping at rate $p(i,j)=p(j,i)$ between $e_i$ and $e_j$.
\el

\bpr
As this is a straightforward extension of Lemma \ref{two}, we will be brief.
On the corner points $\caC$ of the finite simplex $E$ a general function
\be
f(y)=\sum_{i\in S} f(e_i )\, y_i
\ee
can again be written as a linear combination of simple linear functions $y_i$, which implies the simpler martingale characterization along the same arguments as Lemma \ref{two} in the two-site case.
\epr

The generator $L_N =\sum_{i,j\in S} L_N^{ij}$ (\ref{taylorgen}) applied to the linear function $f_k (y)= y_k$ gives
\[
L_N f_k (y)=\sum_{i\in S}\frac{\alpha}{2}p(k,i) (x_i-x_k)
\]
Therefore, for all $k\in S$
\[
y^N_k (t)-y^N_k (0) -\sum_{i\in S}\int_0^t \frac{\alpha}{2}p(k,i) ( y^N_i-y^N_k )
\]
is a martingale. So the limit along any convergent subsequence of $(y^N (t):t>0)$ will fulfill the condition in Lemma \ref{fclem} and is therefore unique. Again, the initial conditions of the right-continuous limit process can be defined as in (\ref{inicon}) and have distribution $\E [\nu_{x^0}]$ by the same arguments,
which proves the second statement of Theorem \ref{theo2}.

\section{The general case\label{sec:gen}}

\subsection{Concentration and tightness\label{sec:tightness}}

Recall the generators \eqref{taylor1} and (\ref{taylorgen}) of the original and the auxiliary process,
\beq
\caL_N f(x)&=& -\sum_{i,j\in S}\frac{\alpha}{4}p(i,j) (x_i-x_j) (\partial_{ij} f) (x)
\nonumber\\
& &+\frac12\sum_{i,j\in S}p(i,j) x_i x_j \theta_N (\partial_{ij}^2 f)(x) + O(\theta_N/N)
\nonumber\\
&=&Lf(x)+\theta_N L'f(x) +O(\theta_N /N)\ .
\eeq
Individual terms for fixed $i,j$ are precisely of the form dealt with in the two-site case.
Therefore, Lemmas \ref{one} and \ref{ttight} on concentration and tightness are relatively straightforward to generalize, and we only give short proofs in the following.

\bl\label{exone}
We have for all $t>0$ for the auxiliary process
\be\label{excl1}
p(i,j)\,\sup_{y\in E}\E_y \big[ y^N_i (t)\, y^N_j (t)\big]\to 0\quad\mbox{as }N\to\infty\ ,
\ee
and furthermore
\be\label{excl2}
p(i,j)\,\limsup_{N\to\infty} \theta_N \sup_{y\in [0,1]}\E_y \big[ y^N_i (t) y^N_j (t)\big] \leq C
\ee
for some $C>0$. The same holds for the original process $\big( x^N(t) :t>0\big)$ with generator $\caL_N$.
\el

\bpr
Fix $i,j\in S$ with $p(i,j)=p(j,i)>0$ (otherwise there is nothing to prove). Analogous to the proof of Lemma \ref{one} we can compute
\be
L' (y_i y_j )=-2 p(i,j) y_i y_j
\ee
and
\be
L (y_i y_j )=-\frac{\alpha}{4} \Big( y_i \sum_{k\in S} p(j,k) (y_j -y_k ) +y_j \sum_{k\in S} p(i,k) (y_i -y_k )\Big)\ .
\ee
The second term is bounded in absolute value by the constant
\[
C:=\alpha\max_{i,j\in S}\sum_{k\in S} (p(i,k)+p(j,k))
\]
and only the first term is relevant, since it is multiplied by $\theta_N$. Using the notation
\be
\phi_t (y)=\E_y \big[ y^N_i (t) y^N_j (t)\big]\quad\mbox{and}\quad \psi_t (y)=\E_y \big[ L (y^N_i y^N_j )\big]
\ee
we get the estimate
\be
0\leq \psi_t (y) \leq  e^{-2p(i,j)\theta_N t}+ \frac{C}{2p(i,j)\theta_N}\left(1- e^{-2p(i,j)\theta_N t}\right)
\ee
with the same arguments that lead to (\ref{clue}). The rest of the proof follows analogously to the two-site case.
\epr

\bl\label{tightness}
The sequences $\big( x^N (t):t>0\big)$ for the full process and $\big( y^N (t):t>0\big)$ for the auxiliary process are tight on the path space $D(0,\infty )$ of rcll functions.
\el

\bpr
We use again the criterion in Lemma \ref{tcrit} on the compact state space $E$ and the same argument as in the two-site case. The only new element is the estimate of the generator \red{which follows analogously using} 
\be
\sup_{y\in E}\E_y \big[ L_N f(y^N (t))\big]\leq \sum_{i,j\in S} \Big( p(i,j) \| \partial_{ij} f\|_\infty +C\| \partial_{ij}^2 f\|_\infty \Big)
\ee
since all $y^N_i (t)\in [0,1]$ and for $t>0$ we can use the estimate \red{Lemma \ref{exone}}. Again, $f$ is smooth on the compact space $E$ so that all partial derivatives are bounded.
\epr

So the limiting process exists for all positive times $t>0$ and concentrates on the absorbing set $\cal A$ of the fast Wright-Fisher diffusion. The initial condition on $\caA$ will be given consistently by right limits of the sample paths which coincide with the absorption probabilities of the Wright-Fisher diffusion, as we will see in the next subsections.

\subsection{Slow-fast Markovian systems and projection of Markov processes\label{sec:projection}}

In the general case, the generator \eqref{taylorgen} still consists of
a fast Wright-Fisher diffusion part and a slow mutation drift part.
The limiting process will take place on the absorbing set $\caA$ of the Wright-Fisher diffusion,
which in the case of general $p(i,j)$ is a richer set than
the set of corner points $\caC =\{ e_i: i\in S\}$.

We first define the limit motion of a general two-scale Markovian system in greater generality, following closely the results in \cite{kurtz}.
Let us consider $\big( X(t):t\geq 0\big)$, $\big( X'(t):t\geq 0\big)$ Markov processes on a compact metric space $E$ with generators $L$, $L'$. Both generators should have a common core $\caK$ (which is given by the smooth functions $C^\infty (E)$ in our case) and the corresponding semigroups $e^{tL}$ and $e^{tL'}$ are defined on the Banach space $C(E,\R )$ of continuous functions with the supremum norm $\| .\|_\infty$. Suppose $X'_t$ has a Borel measurable set of absorbing states $\caA\subset E$ such that the hitting probabilities from any point $x\in E$ are well defined and given by
\be\label{harm}
\nu_x (A):=\lim_{t\to\infty} \pee_x (X'(t)\in A)
\ee
for measurable subsets $A\subset \caA$. Of course we have $\nu_x =\delta_x$ for all $x\in \caA$.

In the spirit of \cite{kurtz}, Theorem 1.11, we define the {\em $X'$-harmonic projection} operator
\be\label{harmex}
P: C(E,\R)\to \caH (E)\quad\mbox{where}\quad  P f(x):= \int_\caA f(a) \nu_x (da)\ .
\ee
for all $f\in C(E,\R)$, where the range of $P$ is the set of $L'$-harmonic functions $\caH (E)$.
Here, $\nu_x (da)$ is the harmonic measure for the generator $L'$ on the set $\caA$ defined in \eqref{harm}. Hence,
the function $Pf$ is harmonic for $L'$ and solves the
Dirichlet problem $L'\psi=0$ with $\psi(a)=f(a)$ for all $a\in \caA$.
In particular $Pf(a)= f(a)$ for $a\in \caA$, and as a consequence $P^2 f= Pf$, i.e.,\
$P$ acts as a projection from the set $C(E,\R)$ onto the set of harmonic functions.

Denote by $S_\kappa (t)$ the semigroup of the two-scale process $X^\kappa$ with generator $L+\kappa L'$. The following result describes the limiting process as $\kappa\to\infty$ by a semigroup on the harmonic functions.

\bt\label{popo}
Suppose that $e^{tL'}f\to Pf$ as $t\to\infty$ for all $f\in C(E,\R )$ and that the operator
\be
A:C(E,\R )\to \caH (E)\quad\mbox{where}\quad A\, f:=PL\, f
\ee
generates a Markov semigroup $S(t)$ on $\caH (E)$ (the range of $P$). Then $S_\kappa (t)f\to S(t)f$ as $\kappa\to\infty$ for all $f\in\caH (E)$ and $t\geq 0$. The limiting process generated by $A$ concentrates on the absorbing set $\caA$ and has initial condition given by the harmonic measure $\nu_{x^*}$ if $X^\kappa (0)\to x^*$ for some $x^* \in E$.
\et

\bpr
Convergence of the semigroups to the semigroup with generator
$A$ follows from \cite{kurtz}, Theorem 2.1. The only condition to check is $\mbox{dom} (A)\subset\overline{\caR (\lambda -A)}$ for some $\lambda >0$.
This condition is clearly satisfied if $A$  generates a Markov process, because in that
case $\caR (\lambda -A)$ equals $C(\caA,\R)$.

With Dyson's formula (\cite{liggett} p.~367) we have for all $f$ in the core $\caK$
\be
S_\kappa (t)\, f-e^{\kappa tL'} f=\int_0^t e^{\kappa (t-s)L'} L\, S_\kappa (s)\, f\, ds\ .
\ee
Using convergence of the semigroups $S_\kappa (t)$ and $e^{\kappa tL'}$ as $\kappa\to\infty$ we get
\be
S(t)\, f-P\, f=\int_0^t A\, S(s)\, f\, ds\ .
\ee
This shows that $A$ generates $S(t)$ and in particular that $S(0)=P$. For harmonic $f\in\caH (E)$ we have $S(0)f=Pf=f$ and $S(t)$ is indeed a semigroup on $\caH (E)$. This implies that the limit process generated by $A$ concentrates on the absorbing set $\caA$ as explained in more detail below, and that by the definition of the projection $P$ (\ref{harmex}) the initial condition of the process is given by the harmonic measure $\nu_x$.
\epr

Every $L'$-harmonic function on $E$ is uniquely determined by its boundary values on $\caA$ (cf.\ (\ref{harmex})). So in addition to the projection $P$ we can define the {\em harmonic extension}
\be
\caP: C(\caA, \R) \to \caH(E)\quad\mbox{where}\quad \caP f(x) := \int \nu_x (da) f(a)\ .
\ee
Notice that $P$ and $\caP$ are defined by the same formula, but act on different spaces. In particular we have $Pf = (\caP (f|_\caA))$, i.e., the harmonic projection is the same as the harmonic extension
of the restriction of $f$ to $\caA$. With this notation we can define
\be\label{projgen}
A':C(\caA ,\R )\to C(\caA ,\R)\quad\mbox{where}\quad A'\, f:=(A\caP\, f)|_\caA \ ,
\ee
which highlights the fact that the limit semigroup generated by $A$ actually defines a process on the absorbing set $\caA$.
Via the identification of $\caH(E)$ with $C(\caA, \R)$ the operators $A$ and $A'$ both
describe infinitesimal motion away from $\caA$ due to $L$, instantaneously followed by projection onto $\caA$.
Formally, in order to define $A'$ on $f\in C(\caA,\R)$ and act with $L$ we first have to extend the function $f$ to the whole
configuration space $E$: therefore we first harmonically extend $f$ (action of $\caP$), then apply $L$, then project (action of $P$) and restrict again to $\caA$.

In order to apply Theorem \ref{popo} we have to assure that the operator $A'$ generates a Markov process. This will be done in the next Section and below, where we explicitly compute $A'$ and recognize it as the generator of a process on $\caA$ with a jump and a diffusion part.
The condition of convergence of the semigroup of the fast diffusive part of the generator in our case is covered by the following result.

\bp
Let $L'=\frac12\sum_{i,j\in S}p(i,j) y_i y_j \partial_{ij}^2$ be the generator (\ref{difgenpartL'}) of the Wright-Fisher diffusion $X'$ with absorbing states $\caA$ (\ref{abss}). Then
\be
\| e^{tL'}\, f-P\, f\|_\infty \to 0\quad\mbox{as }t\to\infty\ .
\ee
\ep

\bpr
Using that $P$ is the projection on the absorbing set we have
\[
(e^{tL'}-P) f(x)=\E_x \big[ f(X'(t))-f(X'(\infty ))\big]\ .
\]
Denoting by $\tau$ the time of absorption of the Wright-Fisher diffusion, we get
\beq
\E_x \big[ f(X'(t))- f(X'(\infty))\big] &=&\E_x \big[\big( f(X'(t))- f(X'(\tau))\big)\, \1 (\tau>t)\big]\nonumber\\
&\leq &2\| f\|_\infty \Pr_x\big[ \tau >t\big].
\eeq
This converges to zero since $f$ is bounded on the compact state space $E$, and the Wright-Fisher diffusion gets absorbed almost surely in finite time with a uniformly bounded mean absorption time (see e.g.\ \cite{huillet2}).
\epr

To illustrate how the projected generator $A'$ of \eqref{projgen} looks like in a concrete case, let us look back at
the example of two sites with $E=[0,1]$, studied in the previous section.
The $X'$-process is the Wright-Fisher diffusion with
generator $L' = 2x(1-x)\partial^2_x$, with absorbing states $0$ and $1$. The $X$ process
is the deterministic process with generator $L=-\frac{\alpha}{2}(2x-1)\partial_x$.
Since $X'(t)$ is a martingale, stopping at the time of absorption gives that
the harmonic measure equals
\[
\nu_x (\{1\})=1-\nu_x (\{0\}) = x\ .
\]
As a consequence, for a function $f:\{0,1\}\to\R$, its harmonic extension is simply given by linear interpolation:
$\caP f(x)= xf(1)+ (1-x) f(0)$ and therefore
\[
(L\caP f)(x)= {-}\frac{\alpha (2x{-}1)}{2} \frac{d}{dx} (xf(1)+ (1-x) f(0)) ={-}\frac{\alpha (2x{-}1)}{2}(f(1)-f(0))\ .
\]
Thus
\[
\red{A'} f(x)=(PL\caP )f(x)=\frac{\alpha}{2}\big[ x(f(0)-f(1))+(1-x)(f(1)-f(0))\big]
\]
which in the case of $\frac{\alpha}{2} =1$ takes values
\[
\red{A'} f(1)= f(0)-f(1)\quad\mbox{and}\quad \red{A'} f(0)= f(1)-f(0)\ .
\]
So $A$ is the generator of the process which flips at rate $1$ between the states of the absorbing set $\caA=\{0,1\}$, as we found in Section \ref{sec:twosite} using a martingale characterization.

\section{Computation of the generator $A$: the case of three sites\label{sec:threesite}}

In the case of general connections $p(i,j)$ and general finite sets $S$, the possible limiting behaviors of the process is more complicated. Therefore we focus on the case of Theorem \ref{theo3} with $p(i,j)\in\{ 0,1\}$ and compute the generator $A$ (\ref{projgen}) using that
\be\label{plp}
PL=P\lim_{h\searrow 0}\left(\frac{ e^{hL} - I}{h}\right) = \lim_{h\searrow 0}P\left(\frac{ e^{hL} - I}{h}\right)
\ee
by continuity of the projection $P$. 
\red{Note that by Theorem \ref{popo} it suffices to compute $A$ rather $A'$ to characterize the limit process.} 
We start here with a system of three sites with closed boundaries and nearest neighbor
connections. This case already contains most of the interesting aspects of the limiting dynamics
which now has both a diffusive and a pure jump part.

\subsection{Generator and absorbing set}
We consider the three site case with $p(1,2)=p(2,1)=p(2,3)=p(3,2)=1$ and $p(1,3)=p(3,1)=0$. The generator (\ref{taylorgen}) is
\beq\label{threegen}
L_N&=& -a (x_1-x_2)\partial_{12} -a (x_2-x_3)\partial_{23}
\nonumber\\
&+&
\frac{\theta_N}{2} x_1 x_2 \partial_{12}^2 + \frac{\theta_N}{2} x_2x_3 \partial^2_{23}
\eeq
where we write $a=\alpha /4$ in the following to simplify notation.
Recall the notation from (\ref{taylorgen}) for this case: the ``fast Wright-Fisher diffusion'' has generator
\be
L'=  \frac{1}{2} x_1 x_2 \partial_{12}^2 + \frac{1}{2} x_2x_3 \partial^2_{23} \ ,
\ee
whereas, the ``slow deterministic'' part has generator
\be\label{detpartthree}
L=-a (x_1-x_2)\partial_{12} -a (x_2-x_3)\partial_{23} \ .
\ee
The process $x'(t)$ with generator
$L'$ is a Wright-Fisher diffusion, for which, $x'_1 (t),\, x'_2 (t),\, x'_3 (t)$ and $x'_1 (t) x'_3(t)$ are  martingales since $L' x_1=L' x_2 =L' x_3=L' (x_1x_3)=0$.
Its absorbing set is
\be\label{3abs}
\caA =(0,z,0)\cup\big\{ (u,0,w):u+w=z\big\} 
\ee
given by a point and a line segment.
Here we assume 
that the total mass in the system is $z\leq 1$, for our results to generalize more easily to larger systems.

\subsection{Dynamics on the line segment: diffusion and jumps}

We start with a point on the line segment $x(0)=(u,0,w)$ and need to compute (see \eqref{plp}):
\be\label{tocomp}
Af(x(0))=\lim_{h\searrow 0}\frac1{h}\Big(\E_{x(h)} \big[ f(x'(\tau  ))\big] -f(u,0,w)\Big)\,
\ee
where $x(h)=\big( u(1-ah),zah,w(1-ah)\big)\in E\setminus\caA$ is the initial condition for the Wright-Fisher diffusion $x'(t)$.
This initial condition comes from starting from the point $(u,0,w)$ from the absorbing set and evolving
according to the deterministic part of the dynamics a small amount of time $h$.

Then, since $x'_2 (\tau  )\in \{ 0,z\}$ takes only two values, we have
\beq
\E_{x(h)} \big[ f(x'(\tau  ))\big] &=&f(0,z,0) \,\Pr_{x(h)} [x'_2 (\tau  )=z ]\nonumber\\
& &+\E_{x(h)} \big[ f(x'(\tau  ))\,\1 (x'_2 (\tau  )=0)\big]\ .
\eeq
Since $x'_2 (t)$ is a martingale on $[0,z]$, the probability in the first line is given by $ah$ and we have
\beq\label{twoparts}
\lefteqn{\E_{x(h)} \big[ f(x'(\tau  ))\big] -f(u,0,w)=ah \big( f(0,z,0)-f(u,0,w)\big)}\nonumber\\
& &+\E_{x(h)} \Big[ \big( f(x'(\tau  ))-f(u,0,w)\big)\,\1 (x'_2 (\tau  )=0)\Big]\ .
\eeq
The first line corresponds to a jump part, where all the mass $z$ ends up in site $2$. The second line has to be a continuous part, since the probability of the event $x'_2 (\tau  )=0$ is of order $1+o(h)$.
Since $x'_1 (t)$ is a martingale we have
\[
\E_{x(h)} \big[ x'_1 (\tau  ){-}u\big] ={-}uah ={-}uah +\E_{x(h)} \big[ (x'_1 (\tau  ){-}u)\,\1 (x'_2 (\tau  )=0)\big]\ ,
\]
where the second equality uses again $\Pr_{x(h)} [x'_2 (\tau  )=z ]=ah$ and the fact that $x'_2 (\tau  )=z$ implies $x'_1 (\tau  )=0$. Therefore
\be\label{mzero}
E_{x(h)} \big[ (x'_1 (\tau  )-u)\,\1 (x'_2 (\tau  )=0)\big] =0
\ee
and similarly for $x'_3 (t)$, and the drift of the continuous part vanishes. It remains to compute the diffusion part.
\beq\label{varcomp}
\lefteqn{\E_{x(h)} \big[ (x'_1 (\tau )-u)^2\big] =\E_{x(h)} \big[ x'_1 (\tau )^2 \big] -2u\E_{x(h)} \big[ x'_1 (\tau ) \big] +u^2 }\nonumber\\
& &=\E_{x(h)} \big[ x'_1 (\tau )(z-x'_2 (\tau )-x'_3 (\tau ))\big] -2u^2 (1-ah) +u^2 \nonumber\\
& &=uz(1-ah) -0-uw(1-ah)^2 -2u^2 (1-ah) +u^2 \nonumber\\
& &=ahuz +O(h^2)\ ,
\eeq
where we have used that $x'_1 (\tau )x'_2 (\tau )=0$ and that $x'_1 (t)x'_3 (t)$ is a martingale. Again splitting the expectation with respect to the value of $x'_2 (\tau )$ we get analogously to the above
\be
\E_{x(h)} \big[ (x'_1 (\tau )-u)^2 \,\1 (x'_2 (\tau  )=0)\big] =ahuz-ahu^2 =ahuw\ .
\ee
The same holds for $x'_3 (t)$, and obviously $E_{x(h)} \big[ x'_2 (\tau  )^k\,\1 (x'_2 (\tau  )=0)\big] =0$ for all $k=1,2\ldots$. To get the covariances we compute
\beq\label{varcomp2}
\lefteqn{\E_{x(h)} \big[ (x'_1 (\tau  ){-}u)(x'_3 (\tau  ){-}w)\,\1 (x'_2 (\tau  )=0)\big] =}\nonumber\\
& &=\E_{x(h)} \big[ x'_1 (\tau )x'_3 (\tau ))\,\1 (x'_2 (\tau  )=0)\big] -uw=-2ahuw\ ,
\eeq
where we used (\ref{mzero}) and that $x'_1 (t)\, x'_3 (t)$ is a martingale. Covariances with $x'_2 (t)$ again vanish.

Now, in (\ref{twoparts}) we have up to second order
\beq
\lefteqn{f(x'(\tau ))-f(u,0,w)=\sum_{i=1}^3 \partial_{x_i} f(x(0))\big( x_i (\tau )-x_i (0)\big)}\nonumber\\
& &+\frac12\sum_{i,j=1}^3 \partial_{x_i}\partial_{x_j}f(x(0))\big( x_i (\tau )-x_i (0)\big)\big( x_j (\tau )-x_j (0)\big)\ ,
\eeq
where higher order terms are of order $o(h)$ and not relevant. Taking expectations, the first order terms vanish, and the second order diffusive terms can be written as (recall that $a=\alpha /4$)
\[
\E_{x(h)} \Big[ \big( f(x'(\tau  ))-f(u,0,w)\big)\,\1 (x'_2 (\tau  )=0)\Big] =\frac12 \alpha huw \partial_{13}^2 f(u,0,v)+o(h)
\]
using the same notation as in (\ref{abbr}). Plugging this and (\ref{twoparts}) in (\ref{tocomp}) we obtain
for the limiting generator $A$:
\be\label{apart2}
Af(u,0,w)=\alpha\big( f(0,z,0)-f(u,0,w)\big) +\frac12 \alpha uw \partial_{13}^2 f(u,0,w)\ ,
\ee
consisting of a jump part and a Wright-Fisher diffusive part with effective diffusivity $\alpha uw$. Notice that as soon as $u$ or $w=0$ the diffusive part vanishes, and the generator consists purely of the jump part where the total mass $z$ moves from site $1$ or $3$ onto site $2$.

\subsection{Dynamics of a single condensate: jump process}

For the other possible initial condition $x(0)=(0,z,0)$ we have $x(h)=\big( zah,z(1-2ah),zah)\big)$ as initial condition for the Wright-Fisher diffusion. Again we have
\beq
\E_{x(h)} \big[ f(x'(\tau  ))\big] &=&f(0,z,0) \Pr_{x(h)} [x'_2 (\tau  )=z ]\nonumber\\
& &+\E_{x(h)} \big[ f(x'(\tau  ))\,\1 (x'_2 (\tau  )=0)\big]\ ,
\eeq
where this time the probability in the first line is of order $1-2ah$. Since $x'_1 (t)\, x'_3 (t)$ is a martingale we have
\be
\E_{x(h)} \big[ x'_1 (\tau )x'_3 (\tau )\big] =(zah)^2
\ee
which implies that the probability to end up in a state $(u,0,w)$ with $u,w>0$ is negligible. Therefore
\be
\E_{x(h)} \big[ f(x'(\tau  ))\,\1 (x'_2 (\tau  )=0)\big] =2ah \Big(\frac12 f(z,0,0)+\frac12 f(0,0,z)\Big)\ ,
\ee
and the generator consists only of jump parts which can be written as
\be\label{apart1}
Af(0,z,0)=a\big( f(z,0,0)-f(0,z,0)\big) +a\big( f(0,0,z)-f(0,z,0)\big)\ .
\ee
Therefore, once all the mass concentrates on a single site this remains, and the corner points $\caC$ are absorbing for the limit process.

The closure of the operator $A$ defined by \eqref{apart1},\eqref{apart2} clearly is the generator
of a well-defined Markov process on the set $\caA$ defined in \eqref{3abs}.
Furthermore, tightness of the sequence of processes generated by $L_N$ (\ref{threegen}) is proved in Lemma \ref{tightness}.
Therefore, we can conclude for the three site case, that the processes generated by \eqref{threegen} indeed
converge to the process generated by the projected generator $A$ defined by \eqref{apart1}, \eqref{apart2}.
This concludes the proof of Theorems \ref{theo2} and \ref{theo3} for the case of three sites.

\section{The general case continued\label{sec:final}}

\subsection{Motion of a single condensate}

First we show that once a point $e_i\in \caC$ is hit in the limit
process, i.e., the situation physically corresponding to a single
condensate, then the process becomes a pure jump process on the corner set
$\caC =\{e_i:i\in S\}$. This in combination with \red{Lemma \ref{tightness}} finishes the proof of the first item in
Theorem \ref{theo2}. Recall the decomposition $L_N =L+\theta_N L'$ of the generator of the auxiliary process given in (\ref{taylorgen}).

\bl
\ben
\item
For every Borel measurable subset $K\subset E\setminus\caC$ we have for the function $\nu_. (K):x\mapsto \nu_x (K)$
\be\label{nul}
(L\nu_. (K)) (e_i )=0\ .
\ee
As a consequence, $Af(x)=0$ for all
$x\in E\setminus\caC$ and hence,
\[
Af (e_i)=\sum_{j\in S} \frac{\alpha}{2} p(i,j) (f(e_j)-f(e_i))
\]
i.e., the projected process is a pure jump process on $\caC$.
\een
\el

\bpr
Fix $K\subset E\setminus\caC$. If $p(i,j)>0$ then starting
from a point $x\in E$ such that $x_k=0, k\not\in \{i,j\}$, the Wright-Fisher diffusion $L'$ remains in the plane $E_{ij}:=\{re_i+s e_j: r,s\in\R\}$ and
hence is absorbed at $e_i$ or $e_j$.
As a consequence
\[
\nu_{e_i +s e_j -se_i}( K)=0
\]
which implies
\[
(\partial_{ij}\nu_. (K))(e_i )=0\ .
\]
Since $K$ was an arbitrary Borel measurable subset of $E\setminus\caC$, we conclude for every
bounded Borel measurable function $f$ with support contained in $E\setminus\caC$ that
\[
\left(\partial_{ij}\int_\caA \nu_. (da) f(a)\right)(e_i )=0\ .
\]
Further, if $i\not\in \{k,l\}$ then obviously, for any smooth function $f: \caA\to\R$
\[
((x_k-x_l) \left(\partial_{kl} \int_\caA \nu_. (da) f(a)\right)(e_i )=0\ ,
\]
because both $x_k$ and $x_l$ are zero when evaluated at $x=e_i$.

Therefore, since $L$ in \eqref{detgenpartL} is a linear combination of terms of the form $-\frac{\alpha}{4}(x_k-x_l) \partial_{kl}$, we conclude that
for any smooth function $f:\caA\to\R$ with  support contained in $E\setminus\caC$
\[
L\left(\int_\caA \nu_x(da) f(a)\right)(e_i )=0\ .
\]
Therefore, we obtain for all $f\in C(\caA ,\R )$
\beq
A f(e_i)&=& L\left( \int_\caA \nu_. (da) f(a)\right)(e_i )=
 L\left(\int_\caC \nu_. (da) f(a)\right)(e_i )=\nonumber\\
 &=&\sum_{j\in S} \frac{\alpha}{2} p(i,j) (f(e_j)-f(e_i))\ ,
\eeq
since $\int_\caC \nu_{e_i} (da) f(a) =f(e_i )$.
\epr

\subsection{The general case of Theorem \ref{theo3}}

The proof of the general case is a combination of the arguments used in the three site case discussed above, and we will be brief in this section avoiding a full treatment with a lot of technical notation. Recall that $p(i,j)\in\{ 0,1\}$ and let $x(0)\in\caA$ be a general initial condition, for which $x_i (0)x_j (0)=0$ whenever $p(i,j)=p(j,i)=1$. Again, under the Wright-Fisher process $(x'(t):t\geq 0)$, with generator $L'= (1/2)\sum_{ij} p(,j) y_iy_j \partial^2_{ij}$, all $x'_i (t)$ and all products $x'_i (t)x'_j (t)$ for which $p(i,j)=0$ are martingales.
Analogously to the computations in Section \ref{sec:threesite}, after the action of the deterministic drift generator $L=\sum_{ij} -a p(i,j) (y_i-y_j)\partial_{ij}$ for infinitesimal time $h$ we have an initial condition $x(h)$ for the diffusion $L'$, and denote by $\tau$ the time of absorption.

The first observation is that at most one initially empty site can gain a macroscopic amount of mass through a jump.

\bl\label{onlyone}
$x_i (0)=0$ and $x'_i (\tau )>0$ is possible for at most one $i\in S$, and in that case $\sum_{j\in S} p(i,j)x'_j (\tau )=0$.
\el
\bpr
Let $i$ and $j$ be two initially empty sites for which $x_i (h), x_j (h)=O(h)$. Then, if $p(i,j)=1$ we have $x'_i (\tau )x'_j (\tau )=0$ since $\tau$ is the hitting time of the absorbing set $\caA$, so that at most one of them can take a positive value. Similarly, if $p(i,j)=0$ we have $x'_i (\tau )x'_j (\tau )=O(h^2)$ since the product is a martingale, and since we take the limit $h\to 0$ again at most one can be positive after the diffusion is absorbed. Since the pair $i,j$ was arbitrary, this implies that under the action of the generator $A$ at most one empty site in the system can gain mass. Then at the time of absorption in $\caA$ all other connected sites have to be empty.
\epr

Secondly, jumps of a macroscopic amount of mass occur only onto neighbouring empty sites.

\bl\label{empty}
If $x_i (0)=0$ and $\sum_{j\in S} p(i,j) x_j (0)=0$, then $x'_i (\tau )=0$. If $x_i (0)>0$ then $x'_i (\tau )\leq x_i (0)+O(h)$.
\el
\bpr
By the condition site $i$ has no direct neighbour with $x_j (0)>0$, and therefore if $x_i (0)=0$ under the action of generator $L$ in time $h$ we get $x_i (h)=O(h^2 )$. Then since $x'_i (t)$ is a martingale, its mass remains negligible after the Wright-Fisher diffusion $L'$ gets absorbed and $x'_i (\tau )=x_i (0)=0$.\\
If $x_i (0)>0$ we also have $\sum_{j\in S} p(i,j) x_j (0)=0$ and thus $x_i (h)\leq x_i (0)+O(h^2 )$. Then the martingale property of $x'_i$ implies that absorption at $x'_i (\tau )=x_i (0)+O(1)$ has negligible probability of order $h^2$.
\epr

Thirdly, mass between two macroscopically occupied sites moves at most continuously.

\bl\label{nointer}
If $x_i (0)x_j (0)>0$, then either $x'_i (\tau )x'_j (\tau )=0$ or $x'_i (\tau )x'_j (\tau )=x_i (0)x_j (0) +O(h)$.
\el
\bpr
$x_i (0)x_j (0)>0$ implies $p(i,j)=0$.
In the first case the mass of one of the sites jumped to another site. Now assume this is not the case and note that $x'_i (t)$, $x'_j (t)$ and $x'_i (t)x'_j (t)$ are martingales.
Let $f,g:[0,1]\to\R$ be two smooth functions, then
\be
L'\big( f(x_i )g(x_j )\big) =\sum_{k\in S}\Big( p(i,k)x_i x_k \partial_{ik}^2 f(x_i )+p(j,k)x_j x_k \partial_{ik}^2 g(x_j )\Big)\ .
\ee
Then $x_i (h)x_j(h)=O(1)$ which implies that $x_k (h)=O(h)$ for all $k$ connected to either $i$ or $j$ and
\be
L'\big( f(x_i (h) )g(x_j (h))\big) =O(h)\ .
\ee
So either a neighbour $x'_k (t)$ gains a macroscopic amount of mass during the Wright-Fisher diffusion, which results in the first case of one of the piles jumping to $k$, or we have
\be
\E_{x(h)} \big[ f(x'_i (\tau) )g(x'_j (\tau)) \big] =f(x_i (0) )g(x_j (0))+O(h)\ .
\ee
Since $f$ and $g$ are arbitrary, the latter case implies the second statement.
\epr

\bl
If $x_i (0)=0$ and $x'_i (\tau )>0$ then $x'_i (\tau )=\sum_{j\in S} p(j,i) x_j (0)+O(h)$.
\el
\bpr
This follows from Lemmas \ref{onlyone} and \ref{nointer}. $i$ is the only site in the system where a macroscopic amount of mass jumped to and its neighbours are empty at absorption, therefore it has to absorb all the original mass of its neighbours up to amounts of order $h$ which have been shared with other sites.
\epr

This characterizes all the possible jump events under the generator $A$. We see that mass can only jump onto a single empty site $j$ which collects all the surrounding mass, and the rate for this event can be computed analogously to the previous section to be $\frac{\alpha}{2}\sum_{i\in S}p(i,j) x_i (0)$. In parallel to jumps, mass can move continuously between macroscopically occupied sites, which we describe next. We will use the notation $\hat p(i,j)$ introduced in (\ref{nnn}) for the number of two-step connections between two sites $i,j\in S$. The crucial ingredient to conclude is additivity of the jump rates of the Wright-Fisher diffusion in each component, as is described below.

\bl
Assume that $\hat p(i,j)x_i (0)x_j (0)>0$ then the generator $A$ contains a term $a\hat p(i,j)x_i (0)x_j (0)\partial_{ij}^2$ and the continuous part of $A$ is given by the sum of all such contributions.
\el

\bpr
For $i,j\in S$ as given, assume that $x'_i (\tau )x'_j (\tau )>0$ so that none of the two piles jumps and the total mass is conserved up to order $h$, which has $O(1)$ probability. Then the system on sites $i,k,j$ with intermediate site $k$ can be seen as an effective three site system as studied in the previous section. If there is a unique intermediate site, the diffusive part of the generator follows analogously to this case. If there are several intermediate sites $k$, the different paths between $i$ and $j$ can be viewed as independent, since the diffusion rates of the Wright-Fisher generator acting on sites $i$ and $k$ given by
\be
L'_{ik}=x_i x_k \partial_{ik}^2
\ee
are proportional to $x_k$ (similarly between $j$ and $k$). Therefore the total effective Wright-Fisher diffusion rate between sites $i$ and $j$ is simply given by a sum over all connections, as given by $\hat p(i,j)$ in (\ref{nnn}). Since the above rates are also linear in $x_i$ (and $x_j$, resp.), the same additivity applies if $i$ or $j$ are connected to more different sites via two-step connections. This leads to independent contributions for each such connection as claimed in the Lemma, which are summed over to give the full continuous part of the generator.

It remains to show that there is no mass exchange over distances more than $2$ steps. This is most easily seen in a system with 4 sites $S=\{ 1,2,3,4\}$ with nearest neighbour jumps and initial condition $(u,0,0,w)$ and $u+w=z$. Analogous to the computation in (\ref{varcomp}) for three sites we get
\[
\E_{x(h)} \big[ (x'_1 (\tau )-u)^2\big] =ah(uz-uw)+O(h^2)\ .
\]
This implies analogously to (\ref{varcomp2}) that
\[
\E_{x(h)} \big[ (x'_1 (\tau )-u)^2 \1 (x'_2 (\tau )=0)\big] =ah(uz-uw)-ahu^2 +O(h^2)=O(h^2)\ ,
\]
so the change of $x_1$ is not on a diffusive scale and negligible. The same holds for $x_4$, and this argument can be easily generalized to arbitrary $S$.
\epr

\section*{Acknowledgements}
S.G.\ acknowledges support by the Engineering and Physical Sciences Research Council (EPSRC), Grant No. EP/I014799/1.

\end{document}